\documentclass[sigconf]{acmart}

\AtBeginDocument{%
  \providecommand\BibTeX{{%
    \normalfont B\kern-0.5em{\scshape i\kern-0.25em b}\kern-0.8em\TeX}}}

\setcopyright{acmcopyright}
\copyrightyear{2021}
\acmYear{2021}
\acmDOI{10.1145/1122445.1122456}

\acmConference[ICSE 2022]{The 44th International Conference on Software Engineering}{May 21–29, 2022}{Pittsburgh, PA, USA}
\acmPrice{15.00}
\acmISBN{978-1-4503-XXXX-X/18/06}

\usepackage{listings}
\usepackage{microtype}
\usepackage{booktabs}
\usepackage{cleveref}
\usepackage{tabularx}
\usepackage{todonotes}
\usepackage{csquotes}
\usepackage{subcaption}
\usepackage{xcolor}
\usepackage{xspace}

\newcommand{\taskonly}[1]{\textsc{#1}}
\newcommand{\Basics}{\taskonly{Basics}\xspace}
\newcommand{\Ownership}{\taskonly{Ownership}\xspace}
\newcommand{\Aliasing}{\taskonly{Aliasing}\xspace}

\newcommand{\task}[2]{\if#2B%
	\textsc{#1}$_{\textit{GC}}$%
	\else
	\textsc{#1}$_{\textit{noGC}}$%
	\fi
}

\DisableLigatures[<]{encoding = T1}
\DisableLigatures[>]{encoding = T1}

\newcommand{\code}[1]{\lstinline|#1|}
\newcommand{\anonymous}[1]{(anonymous for review)}

\crefname{figure}{Figure}{Figures}
\crefname{table}{Table}{Tables}

\copyrightyear{2022} 
\acmYear{2022} 
\setcopyright{rightsretained}
\acmConference[ICSE '22]{44th International Conference on Software Engineering}{May 21--29, 2022}{Pittsburgh, PA, USA}
\acmBooktitle{44th International Conference on Software Engineering (ICSE '22), May 21--29, 2022, Pittsburgh, PA, USA}
\acmPrice{15.00}
\acmDOI{10.1145/3510003.3510107}
\acmISBN{978-1-4503-9221-1/22/05}

\begin{document}

\title{Garbage Collection Makes Rust Easier to Use: A Randomized Controlled Trial of the Bronze Garbage Collector}

\author{Michael Coblenz}
\orcid{0000-0002-9369-4069}
\affiliation{%
  \institution{University of Maryland}
  \streetaddress{8125 Paint Branch Drive}
  \city{College Park}
  \state{Maryland}
  \country{USA}
  \postcode{20742}
}
\email{mcoblenz@umd.edu}

\author{Michelle L. Mazurek}
\orcid{0000-0003-4151-6428}
\affiliation{%
  \institution{University of Maryland}
  \streetaddress{8125 Paint Branch Drive}
  \city{College Park}
  \state{Maryland}
  \country{USA}
  \postcode{20742}
}
\email{mmazurek@cs.umd.edu}

\author{Michael Hicks}
\orcid{0000-0002-2759-9223}
\affiliation{%
  \institution{University of Maryland}
  \streetaddress{8125 Paint Branch Drive}
  \city{College Park}
  \state{Maryland}
  \country{USA}
  \postcode{20742}
}
\email{mwh@cs.umd.edu}


\begin{abstract}
  Rust is a general-purpose programming language that is both type- and memory-safe. Rust does not use a garbage collector, but rather achieves these properties through a sophisticated, but complex, type system. Doing so makes Rust very efficient, but makes Rust relatively hard to learn and use. We designed Bronze, an optional, library-based garbage collector for Rust. To see whether Bronze could make Rust more usable, we conducted a randomized controlled trial with volunteers from a 633-person class, collecting data from 428 students in total. We found that for a task that required managing complex aliasing, Bronze users were more likely to complete the task in the time available, and those who did so required only about a third as much time (4 hours vs. 12 hours). We found no significant difference in total time, even though Bronze users re-did the task without Bronze afterward. Surveys indicated that ownership, borrowing, and lifetimes were primary causes of the challenges that users faced when using Rust.
\end{abstract}



\maketitle

\begin{sloppypar}

\section{Introduction}

Rust is a general-purpose programming language that has an emphasis on performance while also being type-, memory-, and thread-safe~\cite{Rust}. One reason for Rust's efficiency is that it does not use garbage collection (GC). Instead, it imposes a compiler-enforced discipline of \emph{ownership} and \emph{lifetimes} (based on \emph{linear logic}~\cite{Wadler90:Linear} and \emph{region-based memory management}~\cite{tofte97regions,GrossmanMJHWC02}, respectively) to ensure that no object reference will be used after its referent is freed. This discipline imposes restrictions on aliasing: an alias can only be \emph{borrowed} temporarily, and doing so limits mutation (which also helps avoid data races). For example:

\begin{lstlisting}
fn foo() {
  let s1 = String::from("hello");
  let len = calc_len(&s1); //lends reference
  println!("the length of '{}' is {}",s1,len);
  // s1 lifetime ends; dropped
}
fn calc_len(s: &String) -> usize {
  // s.push_str("hi"); <-- not allowed: s immutable
  s.len() // s lifetime ends; but not its referent's
}
\end{lstlisting}
Here, function \code{foo} defines a \code{String} \emph{owned} by variable \code{s1}. It then calls \code{calc_len}  to compute its length by \emph{lending} a reference \code{&s1} to the called function, which may only read it, not write it. When the function returns, the borrowed reference's lifetime ends so it is dropped, which restores full ownership to \code{s1}. When function \code{foo} completes, \code{s1}'s lifetime ends so it is dropped and the data is freed.


%


\subsection{Rust is Hard to Learn and Use}

Despite its performance and safety advantages and a loyal core of devotees~\cite{rustloved}, Rust remains relatively unpopular; the TiOBE index ranks Rust at \#27 as of July 2021~\cite{TiOBE}, and the IEEE Spectrum ranks Rust at \#20~\cite{Spectrum:Rank}. In the interest of bringing the benefits of the design ideas behind Rust to more software projects, it is worth considering why Rust has seen limited adoption.

A partial explanation resides in the difficulty of learning Rust. Fulton et al.~\cite{Fulton2021:Benefits} interviewed and surveyed software practitioners who adopted or attempted to adopt Rust, finding that 59\% of survey respondents felt that Rust was harder to learn than other languages. Seven of the 16 interviewees reported that the biggest challenges in learning Rust were the borrow checker---the part of the compiler that enforces the ownership/borrowing discipline---and the overall change of programming paradigm to one that requires that the compiler be able to reason about lifetimes. Ashley Williams, the interim executive director of the Rust Foundation, agreed: ``[References and borrowing] is notoriously something that people find to be the most difficult part of learning Rust''~\cite{Williams21}. The difficulty of learning Rust has implications on adoption in software teams: 42\% of respondents in Fulton et al.'s survey were concerned about their ability to hire Rust developers, since it would take a long time for new team members to become productive.

In addition to being challenging to learn, Rust can be difficult to use. Programming with DAGs and cyclic data structures is straightforward in most popular languages, but such structures do not conform to Rust's aliasing restrictions. 
Rust provides safe building blocks to work around these restrictions. The structure \code{Rc<T>} is for references to immutable \code{T} objects with manually managed reference counts (confirmed safe by the borrow checker), and \code{RefCell<T>} is a container for a mutable \code{T} object with \emph{dynamically} tracked borrowing. Together (e.g., \code{Rc<RefCell<T>>}) these can implement a discipline of \emph{interior mutability} that supports rich aliasing patterns~\cite{interior-mut}, but in a manner far more complex than most programmers are used to. As such, they may be tempted to use Rust's \code{unsafe} feature to sacrifice safety for ease of use.


\paragraph*{Easing Use with Garbage Collection?}
Rust's restrictions are due to the compiler's inability to verify the safe use of richer, aliased data structures. These restrictions would not be needed if Rust used GC\@. As such, an extension to Rust that included GC could enable programmers to be productive sooner, without having to learn the trickier parts of the language right away. By making the GC \emph{optional}, programmers could still learn and use the harder parts of Rust later, and convert their GC-using code to traditional Rust as needed to improve performance.

Moreover, by making a Rust GC \emph{library-based}, it could even prove useful to experts. While 63\% of the respondents in the survey by Fulton et al.~\cite{Fulton2021:Benefits} cited lack of GC as a reason to use Rust, 87\% cited high performance as a reason---between the two is a category of user open to the idea that high performance and garbage collection are not always at odds. Most code is not performance-critical: a guideline is that 90\% of the time is spent executing only 10\% of the code~\cite{Aho1992:Foundations}. 
Thus GC- and manual-based memory management could coexist. Experiments in Cyclone~\cite{GrossmanMJHWC02}, a C-like systems programming language, found nearly no performance cost of using GC when it was applied judiciously alongside safe, manual techniques~\cite{swamy05experience}.


\subsection{Bronze: A Library-Based GC for Rust}

This paper presents Bronze,\footnote{So named for bronze's corrosion resistance.} a new Rust library that provides a clean garbage collection interface, and an experiment evaluating the possible benefits of using Bronze while learning Rust. By creating a library in which GC is \emph{optional}, we were able to study the effects of GC without including the rest of the language design as an independent variable in the study.

Bronze provides a structure, \code{GcRef<T>}, that implements a garbage-collected reference to a mutable object; such references are not subject to Rust's aliasing restrictions. Bronze uses LLVM stack maps to automatically find roots, obviating the need for programmers to specify tracing roots manually, as is necessary in some existing Rust GCs~\cite{Josephine, Shifgrethor}.

We deployed Bronze in an IRB-approved, randomized, controlled experiment in a sophomore-level programming languages course. Because the course was required for graduation at a university with a large computer science enrollment, we were able to recruit from a population of 633 students who were enrolled in the course. All students carried out a multi-part Rust programming assignment, but those who agreed to participate in the research were randomly assigned to condition \emph{Bronze} or \emph{Traditional}; the former group used Bronze in the assignment while the latter group did not. Ultimately, 333 students were part of the random assignment, and 428 students participated in the study in some way.

The experiment design is shown in \cref{tab:experiment-design}; the tasks are described in detail in \cref{ssec:tasks}. In the assignment, students were given functions and declarations that needed to be completed; Bronze participants were given versions of the interfaces that were structurally similar but which had been adapted to use GC\@.  The assignment parts were cumulative, and the final step for Bronze participants was to redo their previous implementation \emph{without} using GC, ensuring all participants learned how to use traditional Rust. We provided students with unit tests (including source code), and we assigned grades according to which of the unit tests passed. 

We found, after conducting the experiment, that Bronze's interface required small changes to ensure soundness. Because of the small size of the required changes, which are described in more detail in \cref{sec:limitations}, we believe that our results apply to the revised version of Bronze as well.

\renewcommand{\arraystretch}{1.1}
\begin{table}
\begin{tabularx}{\columnwidth}{l X X}
\toprule
\textbf{Topic} & \textbf{Traditional task} & \textbf{Bronze task} \\
\midrule
Basics & \task{Basics}{T}  & \task{Basics}{T}  \\
Ownership, lifetimes & \task{Ownership}{T}  & \task{Ownership}{B} \\
Aliased, mutable data & \task{Aliasing}{T} & \task{Aliasing}{B} \\
Aliased, mutable data	& (none) & \task{Aliasing}{T} \\
\bottomrule
\end{tabularx}
\caption{Tasks in each condition. Subscripts indicate versions of tasks adapted to use GC, or not. Bronze participants completed \task{Aliasing}{T} after completing \task{Aliasing}{B}.}
\label{tab:experiment-design}
\end{table}

\paragraph*{Study Results}
At the conclusion of the study we carried out both quantitative and qualitative analysis of measured data (e.g., project score) and survey responses. Our large sample enabled us to reserve 10\% of the participants' data (with their survey responses) for exploratory analysis. This allowed us to explore which hypotheses might be worth testing while preserving soundness, since the results in \cref{sec:results} are based on the remaining 90\% of the data.

We did not observe a difference between conditions in \emph{completion rates} (fraction of students scoring 100\%) or in \emph{times} completing \Ownership. 
However, students who used Bronze were more likely to complete their aliasing task (\task{Aliasing}{B}) than Traditional students were to complete theirs (\task{Aliasing}{T}), and did so significantly \emph{faster}, spending a median of 4 hours instead of 12 hours. The additional time spent by Bronze students doing the aliasing task again, this time without GC (\task{Aliasing}{T}) resulted in no significant difference in \emph{total time} spent on the project. We conclude that for newcomers to Rust, if the goal is simply to accomplish a programming task, garbage collection may present a significant benefit for productivity. Further, there may be enough advantage to using garbage collection while learning Rust to compensate for the additional time required to learn, apply, and switch to traditional Rust memory management approaches.

Survey results confirmed past reports~\cite{Williams21, Fulton2021:Benefits} that ownership and borrowing are significant programming challenges. Participants were much more likely to believe that GC makes writing programs easier after completing the experiment than they were before the experiment. Bronze participants were more likely to believe that GC made writing programs easier after completing the whole assignment than they did at the beginning, and 64 of the 89 Bronze participants (72\%) who responded to the question strongly agreed at the end that using GC makes writing programs easier. Only 3 (3\%) strongly disagreed.
Interestingly, although participants reported that GC made programming easier, participants who used GC did not report \emph{liking} Rust significantly more than participants who did not. We observed a stronger negative correlation between liking Rust and stress than between liking Rust and time spent relative to participants' expectations.

\begin{figure*}
\begin{subfigure}[t]{.49\textwidth}
  \begin{lstlisting}[numbers=left, xleftmargin=9pt]
  pub struct IntContainer { n: i32 }
  
  pub fn set(c: &Rc<RefCell<IntContainer>>, n: i32) {
    let mut m = c.borrow_mut();
    m.n = n;
  }
  
  pub fn make_two_references() {
    let c1 = Rc::new(RefCell::new(IntContainer{n: 42}));
    let c2 = c1.clone();
    // c1 and c2 both reference the same object.
  
    set(&c2, 42);
    set(&c1, 43);
    // Now both reference an object with value 43.
  }
  \end{lstlisting}
\subcaption{Two mutable references to a value using interior mutability.}
\label{fig:interior-mutability}
\end{subfigure}
\hfill
\begin{subfigure}[t]{.49\textwidth}
  \begin{lstlisting}[numbers=left, xleftmargin=9pt]
  #[derive(Trace, Finalize)]
  pub struct IntContainer { n: i32 }
  
  pub fn set(mut c: GcRef<IntContainer>, n: i32) {
    c.n = n;
  }
  
  pub fn make_two_references() {
    let c1 = GcRef::new(IntContainer{n: 42});
    let c2 = c1;
    // c1 and c2 both reference the same object.
  
    set(c2, 42);
    set(c1, 43);
    // Now both reference an object with value 43.
  }
  \end{lstlisting}
  \subcaption{Two mutable references to a value using Bronze.}
\end{subfigure}
\caption{A comparison of mutable aliasing with and without Bronze. The interior mutability version (left) requires manually borrowing a mutable reference to the contents of the \code{RefCell} (line 4). Then, the reference count must be manually incremented via \code{clone} (line 10). With Bronze (right), no borrowing is needed (line 5) and a second reference can be obtained with plain assignment (line 10). The \code{Trace} and \code{Finalize} traits needed for GCed objects can be derived automatically (line 1).}
\label{bronze-comparison}
\end{figure*}

\subsection{Implications}

Software engineers should be aware of a possible productivity benefit of garbage collectors relative to using Rust's aliasing restrictions; it may be better to re-architect a system, use a garbage collection library, or use a garbage-collected language if the architecture cannot be changed. If engineers use a library-based GC with Rust and need to remove it to improve performance, using GC saves enough time that programmers can switch away from GC without a significant loss in productivity. 

That positive feelings about Rust were more strongly correlated with frustration and stress than with time spent on the assignment suggests that language designers who want to promote adoption (by making languages programmers \emph{like}) should consider focusing on how to reduce stress, such as by making progress more predictable, rather than how to (only) maximize programmer productivity.

In survey responses, participants reported extreme difficulty with references, lifetimes, and ownership. Participants said \emph{examples} and \emph{live coding} demonstrations helped them learn these concepts most effectively. These responses have implications for pedagogy: we hypothesize that using examples and live coding demonstrations is more effective to explain these challenging Rust concepts than traditional slide-based explanations.

\subsection{Contributions}
  
Key contributions of this paper include:

\begin{enumerate}
	\item The design and prototype implementation of Bronze, a GC for Rust that is simpler to use than prior Rust GCs.
	\item A randomized controlled trial of Bronze, in which we found that Bronze can enable more people to complete tasks within time limits and, among those who finished, significantly reduce time required. We also collected qualitative data, confirming that ownership, lifetimes, and references are particularly challenging for new Rust programmers.
\end{enumerate}

\section{Bronze: Design and Implementation}
\label{sec:bronze}

Bronze introduces \code{GcRef<T>}, which represents a reference to a value of type T that exists in a garbage-collected portion of the heap. \code{GcRef<T>} implements the \code{Deref} trait, so the \code{*} operator can be used to obtain a reference to the underlying value. If one has a mutable \code{GcRef<T>}, the reference can be used to mutate the value. \code{GcRef::new(v)} moves value \code{v} into the garbage-collected portion of the heap and generates a \code{GcRef} pointing to it. 

Rust permits only one mutable reference to a value at a time. For greater flexibility, (standard) Rust supports \emph{interior mutability} of an object through an immutable reference to it~\cite{interior-mut}. The programmer may borrow a special reference to the value that permits mutation, and dynamic checks ensure that only one such reference can exist at a time, enabling a safe relaxation of compile-time checks. With Bronze, mutation is permitted through \emph{all} references to each garbage-collected object, with no extra effort. For example, \cref{bronze-comparison} shows how \code{GcRef} simplifies code when there are multiple mutable aliases. Bronze does not guarantee thread safety; as in other garbage-collected languages, it is the programmer's responsibility to ensure safety.

Bronze is a precise, mark/sweep garbage collector. We selected this design rather than using a conservative collector~\cite{boehm97gc} because precise collection has the potential for better performance and complete collection of garbage. As our primary objective in the design was usability, we designed Bronze to find roots automatically. Prior garbage collectors for Rust either include \code{root} and \code{unroot} methods that must be manually called by the user of the GC library~\cite{Josephine, Shifgrethor} or require references to be cloned manually~\cite{rust-gc, Shredder}.

Bronze defines the \code{Trace} trait, which indicates functionality used by the garbage collector to trace through the object graph to find live objects. Only types that implement \code{Trace} can be put on the garbage-collected heap. 
Bronze also defines the \code{Finalize} trait, which allows users to write code that runs just before an object is deallocated by the collector; it serves as an alternative to \code{drop}, which is for deallocation that runs at a statically-determined time.
Bronze provides macros that automatically derive implementations of \code{Trace} and \code{Finalize} for straightforward types, but users can provide their own implementations if needed.

Rust's compiler translates Rust code to LLVM IR, which has a primitive that allows emitting \emph{stack maps} to annotate the stack with compiler-specified metadata. In the case of Bronze, the metadata allow the runtime to determine which stack addresses correspond with objects that must be traced because they may reference garbage collected objects. The Bronze tracer relies on a modified version of the Rust compiler that emits stack map information in the emitted LLVM IR. The particular stack map mechanism used by Bronze assumes that the program is single-threaded; other available mechanisms could relax this requirement in the future at additional engineering cost for Bronze.

Bronze uses a mark/sweep algorithm based on Goregaokar's implementation~\cite{rust-gc}. To identify which objects can be collected, the runtime keeps a linked list of all objects that it allocated. Then it can collect objects on the list that were not marked by the tracer. 
Bronze's mark/sweep implementation is only a proof of concept, however, and is not production-ready. In particular, the implementation can trace local variables of type \code{GcRef<T>}, but additional work is required to enable tracing of arbitrary types. In the experiment, we used a version of Bronze that \emph{never collects}. This approach was suitable for the programs in the experiment, since they do not allocate enough memory to require collection. We think it is unlikely that a full, performant implementation would require changes to the design, since the work that remains should be beneath the interface.

While this paper was under review, we found that \code{GcRef}'s \code{as_ref} and \code{as_mut} methods could be abused to create multiple mutable \code{&} references to GC objects, breaking Rust’s invariant of allowing only one mutable reference to an object. The \code{Deref} trait implementation on \code{GcRef} could be abused in the same way. \Cref{sec:limitations} describes how we obtained soundness in a revised version of Bronze and why we believe the changes, although slightly reducing usability, do not have a significant impact on our results.

Why is Bronze easier to use than interior mutability? We believe that the primary cause is that Bronze allows free persistent (in-field) aliasing. These aliases can be obtained without using a separate data structure (e.g., \code{Rc}) and without invoking a special method to obtain them (e.g., \code{borrow}). When references need to be passed in fields of structures, the structures can be initialized and managed without explicitly manipulating the lifetimes of the references. In addition, \code{GcRef<T>} provides a unified interface compared to \code{Rc<RefCell<T>>}, which requires understanding two separate structures and how they can be used together. The distinction between these structures can appear blurry to users due to Rust's ``Deref coercion'' feature, which means that programmers can omit \code{*} when dereferencing in certain (but not all) cases. Finally, garbage collection works even in the presence of cycles, which must be managed manually when reference counting is used.

Bronze is available on Cargo, the Rust package manager.\footnote{https://crates.io/crates/bronze\_gc}

\section{Method}

We conducted an experiment in which we randomly assigned participants to use either traditional Rust or Bronze when completing a multi-part programming assignment. We measured their performance and surveyed them about their experiences. Complete task and survey materials are available in the artifact that accompanies this paper.\footnote{https://doi.org/10.5281/zenodo.6045904} The study was approved by our IRB. 

\subsection{Tasks}
\label{ssec:tasks}

We devised a multi-part programming assignment for our study. The specification of each part differed slightly depending on whether participants should use GC or traditional Rust; we label a task with subscript \textsc{GC} or \textsc{noGC} to clarify this, as needed.

The \Basics task, carried out using traditional Rust, focused on the basics of Rust syntax.  The \Ownership task introduced ownership and borrowing. This and later parts involve programming a simulation of turtles living on the university campus. The instructions\footnote{Instructions have been slightly edited for space.} included:

\begin{lstlisting}
In turtle.rs, implement:
* `new` function according to the given signature. 
* Accessors `walking_speed`, `favorite_flavor`, 
  `favorite_color`, and `name`. 

Campus should maintain a vector  (`Vec`) of turtles. 
  In campus.rs, implement methods:
* `new`: creates a new, empty Campus
* `size`: returns the number of turtles on campus
* `add_turtle`: adds a new Turtle to campus.
* `get_turtle`: returns a reference to an turtle at a 
   given index.
* `turtles`: returns an iterator that a caller can use to 
   iterate through the turtles.
* `fastest_walker`: Returns None if the campus is empty. 
   Otherwise, returns Some of a reference to the turtle 
   with the fastest walking speed.
* `breed_turtles`, which uses the functions in `genetics` 
  to breed two turtles, resulting in a new Turtle.(*@\ldots@*)

  Every turtle has a name. Of course, as with people, 
  several turtles may have the same name. In campus.rs, 
  implement `turtles_with_name` so that it returns a 
  vector of turtles that have the given name. 
\end{lstlisting}

Completing this part required understanding ownership transfer (\code{add\_turtle}), references (\code{get\_turtle}), iterators (\code{turtles}), options and borrowing (\code{fastest\_walker}), and mutable vectors (\code{add\_turtle} and \code{breed\_turtles}).  In \task{Ownership}{T}, the \code{get\_turtle} method of \code{Campus} returned a value of type \code{TurtleRef<'a>}, which was defined as \code{\&'a Turtle}. In \task{Ownership}{B}, \code{TurtleRef} was defined as \code{GcRef<Turtle>}.

The \Aliasing task introduced the requirement of multiple mutable references. The instructions included:

\begin{lstlisting}
Change `Turtle` so that each turtle has a field that 
keeps track of its children in a `Vec`. `Vec` of what? 
You'll have to figure that out. Do NOT store indices 
into the Campus's vector because those may change in 
the future (some day we may support removing turtles 
from Campus). Do NOT invent your own indexing scheme 
and store a map somewhere.

If you breed two turtles, each parent should include 
the child in its vector of children. In turtle.rs, 
implement methods:
* `num_children`
* `teach_children`

Note that this task may require you to revisit some of 
the decisions you made in Basics. You are welcome to 
copy/paste implementations from your Basics work, but 
note that some of the signatures are different (in order 
to facilitate the design changes you will need to make).

Your previous implementation of `turtles_with_name` had 
to do a linear search through the whole vector of
turtles(*@\ldots@*) Improve the performance of `turtles_with_
name` by adding a cache to Campus(*@\ldots@*)

\end{lstlisting}

\code{Campus} needed a vector of turtles, each of which needed a vector that had references to the same turtles referenced by \code{Campus}. Because of \code{breed\_turtle}, the turtles needed to be mutable. Finally, the cache required returning collections that referenced the same turtles. 

For \task{Aliasing}{T}, these requirements could be addressed using reference counting and interior mutability. The \code{get\_turtle} method of \code{Campus}, as in \task{Ownership}, returned a value of type \code{TurtleRef}, but \code{TurtleRef} was now a struct with fields that the participants needed to define. \code{TurtleRef}, in turn, exposed a method \code{borrow\_turtle}, which returned a value of type \code{BorrowedTurtle}. \code{BorrowedTurtle} implemented the \code{Deref} and \code{DerefMut} traits, allowing clients to obtain a temporary mutable reference. Participants were required to fill in the fields of \code{TurtleRef} and \code{BorrowedTurtle}; a typical solution was \code{Rc<RefCell<Turtle>>} and \code{RefMut<'a, Turtle>}. This approach allowed external clients to obtain references to \code{Turtle}s that could be made temporarily mutable if needed for the application.

In \task{Aliasing}{B}, \code{TurtleRef} is defined as \code{GcRef<Turtle>}, just as in \task{Ownership}{B}---because \code{GcRef} supports mutation of the referenced value, there is no need for additional structure.

Bronze participants were asked to complete task \task{Aliasing}{T} after completing \task{Aliasing}{B}.

\subsection{Recruitment}

We recruited participants from the 633 students who were enrolled in a required, sophomore-level programming course. Our programming assignment was part of the course's grade, but participation in the research was voluntary, and confirmed by informed consent. Research participants were randomly assigned to use either Bronze or traditional Rust, and agreed to take a survey after completing each part of the assignment. In doing so, they received extra credit on the assignment---1\% extra credit per survey, or 5\% for all three. Participants were free to withdraw from the experiment at any time; students who started the experiment but withdrew received 2.5\% extra credit. Students who withdrew, or opted not to participate at all, could complete any version of the assignment. 
Students had the option of accepting random assignment and/or carrying out surveys but \emph{not} having their data included in our research analysis; we awarded extra credit independently of this choice. Only three students did not consent to their data being used.\footnote{Students were required to opt out if they were under 18 years old.}

\subsection{Procedure}

The instructor gave lectures on programming in Rust during four 80-minute class periods over a period of two weeks (April 13--27). Lecture slides can be found in the paper supplement. The \Basics task in the assignment overlapped with, and was due at the end of, the second week of lectures (April 21--29). The remaining parts of the assignment were released on April 29 and were due May 11.

The README file for the \Ownership and \Aliasing tasks described the study in detail and linked to a Qualtrics survey, which included a consent form and requested demographic information from students who consented. The form also asked for their university ID number so that we could associate student grades with participants. The Qualtrics tool randomly assigned participants to a condition (Bronze or Traditional) upon consent, emailing them which condition they were assigned to. The email also contained a personalized link to a survey to fill out after completing each part of the programming assignment, allowing us to track which students had completed the surveys.

The course used a question-and-answer web site, Piazza~\cite{Piazza}, to allow the students to ask questions about course content. Because we had revised the Rust content relative to the prior semester, the first author joined the teaching assistants (TAs) in answering questions about the assignment. We made sure to answer all questions in a timely fashion with high quality. In addition, as a result of posts on Piazza indicating that students found the assignment difficult, the first author conducted live-coding demonstrations on May 6 and 7.\footnote{Student privacy regulations preclude us from sharing the videos of these demos.} Demonstrations delved into topics covered in class, including \code{Rc}, \code{RefCell}, smart pointers, mutability, scope and borrowing, a comparison between GC and reference counting, string literals, interior mutability, mutable structs, lifetime specifiers, and \code{Box}. The Piazza forum and live demonstrations were not separated by experimental condition, so these did not provide additional data about perceived difficulty.

\section{Limitations}
\label{sec:limitations}

Students' abilities to complete the assignment depended to some extent on the quality of our instruction and the extent to which we emphasized each topic. To help decouple our instructional design from the experiment, we based our instructional materials on the online Rust book~\cite{Klabnik2018:Rust} and leveraged materials that had been used successfully in previous editions of the course.

Because the first author taught the live coding demonstrations and answered many student questions on Piazza, it is possible that this could have introduced bias. One of two course instructors is also a co-author, so there could have been bias in teaching as well, since GC and interior mutability, which were taught in the course, were of interest in the study.

Our study focused on programmers who are new to Rust; our sample was of students. While professionals (many of whom are also new at Rust) would have more programming experience, prior work ~\cite{Naiakshina2020:Conducting, Naiakshina2019:If, Acar2017:Security} suggests that results with students corresponds with results with professionals in related, but not identical, tasks.

All times spent by participants were self-reported. Although we asked participants to use a tool to track time spent for \Ownership and \Aliasing, we did not confirm that they did so. However, any noise or bias in reported times are likely to be consistent across experimental conditions. Although we asked participants to complete the surveys immediately after completing each part of the assignment, some participants submitted the surveys in batches.

The extra credit incentive could have resulted in a non-uniformly selected sample. However, as we discuss in more detail in \cref{sec:discussion}, the small difference in median class grade between participants and non-participants (90\% vs. 87\%) suggests that this does not result in a significant threat to validity. 

More Bronze participants withdrew from the study than Traditional participants (see \cref{sec:discussion}); the withdrawing Bronze participants may have been weaker, less industrious, or more risk-averse than the students that remained, leaving a stronger overall Bronze population compared to the traditional one. We believe the magnitude of the difference in times between conditions (\cref{sec:times}) is large enough that it cannot be explained by this possibility.

All of the students worked on the assignment in the same timeframe, so trends over time could have been due to external influences, such as stress caused by the approach of final exams. However, these trends would have been equally applicable across experimental conditions.

While this paper was under review, we found that \code{GCRef} was unsound for the reasons discussed in \cref{sec:bronze}. To make Bronze sound, we removed the \code{Deref} implementation, replacing it with new \code{borrow} and \code{borrow_mut} methods. These methods return objects whose lifetimes are tracked dynamically, similar to \code{RefCell}'s borrowing mechanism. We also removed the \code{as_ref} and \code{as_mut} methods, which were only used in one of the student submissions.

There is a risk that had the students used the corrected version of \code{GCRef}, they might not have seen the same benefits. To understand whether this might be so, we applied the corrected version of Bronze to a sample of student implementations; we believe that our results would largely hold up, even if students had used this version. We manually modified student programs until we were confident we had seen the relevant challenges; in doing so, we modified 43 student submissions. Two changes were common: inserting calls to \code{borrow} or \code{borrow_mut}; and introducing local variables to extend lifetimes of borrows. In the vast majority of cases, calling \code{borrow} or \code{borrow_mut} sufficed; this could be done automatically by a compiler, similar to \code{Deref} coercion, which is already supported by the compiler. Some cases required manipulating lifetimes with additional local variables or inserting curly braces, which is a common technique in Rust. Rust programmers need to be able to do this already (indeed, we taught this technique in our lectures). Two programs manipulated in this way panicked due to multiple borrows after the naive transformation above; both of these used \code{match} on something that was borrowed, and then took a mutable reference inside one of the cases. A small refactoring addressed these. One case that used an iterator required inserting a \code{*} operator to dereference the iterator's returned reference. 

\section{Results}
\label{sec:results}
In this section, we describe the analysis we conducted. Anonymous raw data as well as an RStudio script to analyze the data are included in the artifact that accompanies the paper.\footnote{https://doi.org/10.5281/zenodo.6045904}

\subsection{Participants}

Of 633 students who were enrolled in the course, 385 students signed up for the experiment, with 190 assigned to use Bronze and 195 to Traditional. Of these, 41 withdrew; an additional 11 students submitted code for both versions of the assignment. We did not analyze the data from those 52 students, leaving experimental data from 333 students for analysis: 139 Bronze, and 194 Traditional.

383 students completed 1120 surveys. 34 students submitted surveys but did not accept random assignment. 36 accepted random assignment but did not submit surveys. Overall, 428 students participated in the study. Only the data from the 333 experiment participants were used in the analyses below, except for the overall demographics of the population and the analysis of whether GC participants thought GC makes programming easier (\cref{sec:gc-easier}). For each task, we analyzed the corresponding survey data, as shown in \cref{tab:survey-submissions}.

\renewcommand{\arraystretch}{1}
\begin{table}
\begin{tabular}{l l l}
\toprule
\textbf{Component} & \textbf{Traditional} & \textbf{Bronze}\\
\midrule
Random assignment to condition & 194 & 139 \\
\Basics survey & 154 & 117 \\
\Ownership survey & 153 & 113 \\
\Aliasing survey & 142 & 101 \\
\task{\Aliasing}{B} survey & & 84 \\
\bottomrule
\end{tabular}
\caption{Participation by component}
\label{tab:survey-submissions}
\end{table}

We asked students about their Rust experience. Of the 333 experiment participants, 84 had read about Rust or talked to a friend about it. 12 had played with Rust on their own, two had used it on a team, and three had used it in an open-source project. When asked to self-assess their prior level of Rust knowledge, 307 reported no experience; 17 reported passing familiarity (maybe wrote a few lines of code); 6 reported a moderate amount or a lot of experience.

\begin{figure}
\includegraphics[width=\columnwidth]{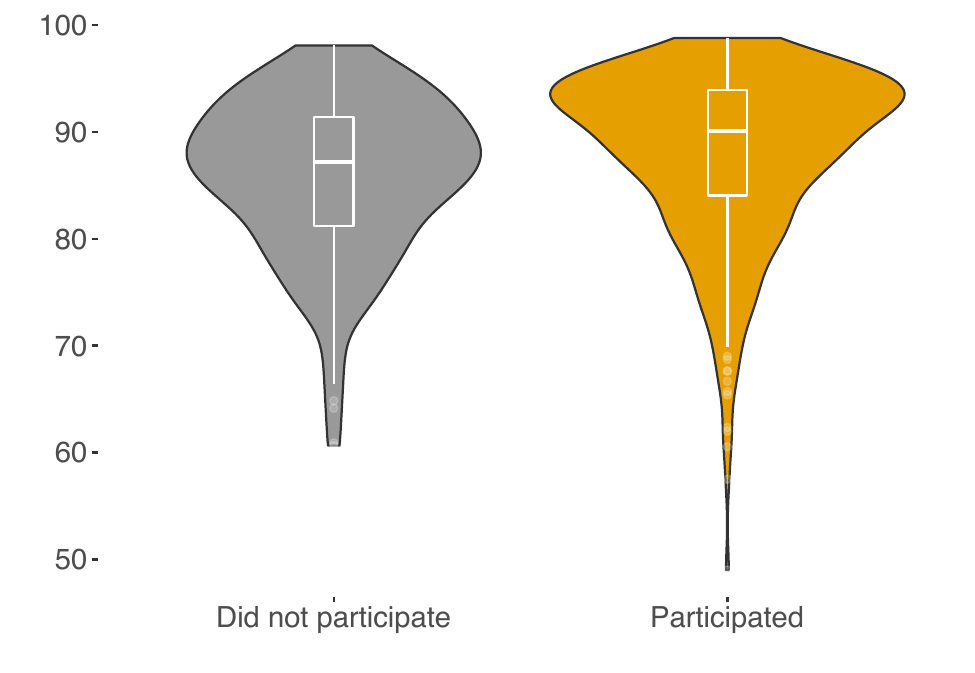}
\caption{Course grade by experiment enrollment}
\label{fig:enrolled.boxplot}
\end{figure}

\Cref{fig:enrolled.boxplot} shows the distribution of course grades by decision to enroll in the experiment in superimposed violin and box plots. To test whether the decision to enroll was related to course grade, we conducted a logistic regression. We found $p \approx .017$, with higher grades being associated with choosing to enroll in the study. The odds ratio was $0.035$ with a 95\% confidence interval of $[0.011, 0.057]$ ($N = 584$), meaning that a 1-point increase in grade correlates with multiplying the odds of enrolling by $e^{.035} \approx 1.036$.

\subsection{Analysis Methods}
We designed the overall experiment to allow us to examine the relationship between time spent and outcomes (grades and completion);  the association between testing practices, sources of help, and editor selection and outcomes; and the effect of condition on the amount of time and fun students predicted if they had done the assignment in \emph{other} languages, as well as the effect of condition on amount and sources of help.

In order to help us refine our list of hypotheses while ensuring a sound data analysis, we randomly reserved data from 10\% of the participants for exploratory analysis. We visualized this exploratory data and conducted statistical tests to identify interesting hypotheses about the effects of garbage collection and other factors that might influence outcomes. From this exploration, we selected specific questions of interest and associated hypothesis tests. Finally, we discarded the reserved 10\% and performed the planned tests on the remaining 90\%. All results reported below are drawn from this 90\%, except overall demographic information (e.g., participation rates), which is reported for the entire dataset. 

Because we conducted multiple hypothesis tests, we interpret the results with a Holm-Bonferroni correction~\cite{Holm1979:Simple}. All of the reported p-values are \emph{corrected} and can be compared directly with $\alpha$; we set $\alpha = 0.05$ in our interpretation.

\subsection{Completion rates}

\begin{figure}
\includegraphics[width=\columnwidth]{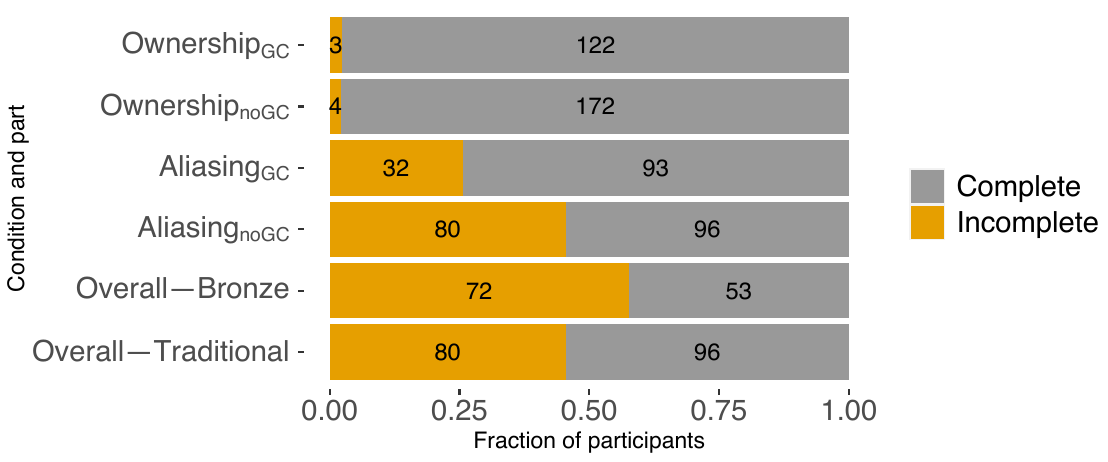}
\caption{Fractions of participants who completed each part}
\label{fig:all.fractions.completed}
\end{figure}

We wanted to know whether Bronze participants were more likely to finish either individual parts of the assignment or the assignment as a whole. We conducted Fisher's exact tests to compare completion rates (rates of scoring 100\%) on each part of the assignment across conditions. This analysis concerns 301 of the 333 total experiment participants, which omits the 10\% of the participants that were withheld for exploratory analysis. Completion rates of each part are shown in \cref{fig:all.fractions.completed}. For \Ownership, we found no significant difference in completion rate ($p \approx 1.00$). Comparing \task{Aliasing}{T} (Traditional) and \task{Aliasing}{B}, we found that Bronze users were significantly more likely to score 100\% ($p \approx .006$). The odds ratio was approximately $0.41$, indicating that the odds of finishing for participants \emph{not} using Bronze were approximately $0.41$ times the odds of finishing for participants who DID use Bronze. When considering probability of completing all parts, Fisher's exact test indicated no significant difference between conditions ($p \approx .51$).

\subsection{Completion times}
\label{sec:times}

\begin{figure}
\includegraphics[width=\columnwidth]{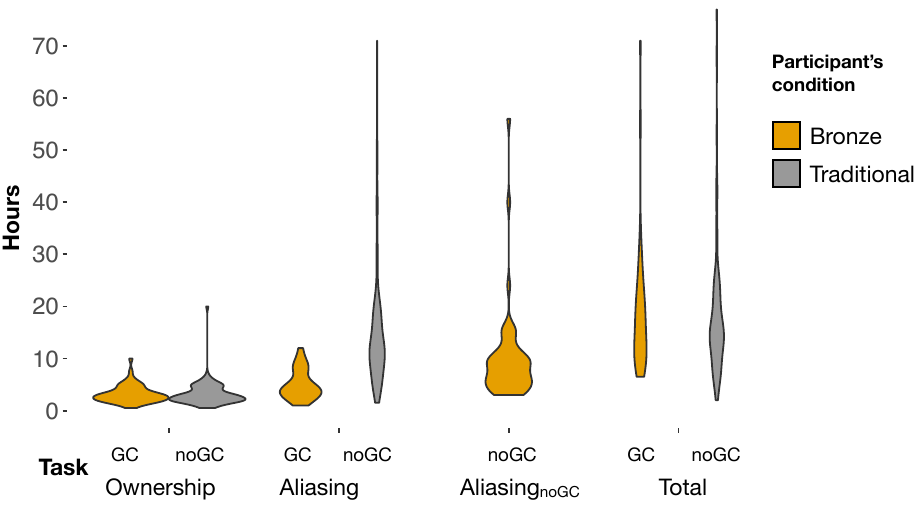}
\caption{Total reported time spent on parts after \Ownership.}
\label{fig:time.boxplot}
\end{figure}

For tasks after Basics (i.e., after participants were assigned to conditions), we hypothesized that Bronze would enable participants to complete tasks \emph{faster}. We analyzed the 149 participants who finished all the tasks. For these tasks, a Shapiro-Welk normality test found a likely violation of normality of the completion times ($p < .001$). 
Therefore, we conducted nonparametric Wilcoxon tests rather than ANOVA tests. \Cref{fig:time.boxplot} shows the total times reported across the conditions. \Cref{fig:time.boxplot} excludes the Basics task, since that task preceded assignment to conditions and because we only asked participants to start tracking their time \textit{after} the Basics task was complete. Nevertheless, we asked participants to estimate how long they spent on the Basics task, and they reported a mean of 3.5 hours ($\mathrm{SD} \approx 3$ hours). Task \task{Aliasing}{noGC} is not compared to any other task; instead, we leverage it for computing the total time spent on all tasks (column ``Total'' in \Cref{fig:time.boxplot}).

We did not find a significant difference in \Ownership time ($W = 2449.5, p \approx 1$). However, Bronze participants finished \Aliasing significantly faster (median = 4 hours) than Traditional Rust participants (median = 12 hours) ($W = 561, p < .001$). We did not find a significant difference in total completion time ($W = 2354.5, p \approx 1$). 

We also compared completion times spent by participants who did \emph{not} finish. A Wilcoxon rank sum test found no significant difference ($W = 829, p \approx 1$). We also compared times reported by participants who scored 100\% to times by participants who did not, finding no significant difference ($W = 5675, p \approx 1$). 

\Cref{fig:time.boxplot} also shows the distribution of times for \task{Aliasing}{T} when completed by Bronze participants. The median time was 8 hours (mean = 9.9, $SD = 8.4$). The difference between medians of \Aliasing times across conditions was also 8 hours, explaining why we did not find a difference in total time spent across conditions.

\subsection{Does GC make writing programs easier?}
\label{sec:gc-easier}
\begin{figure}
\includegraphics[width=\columnwidth]{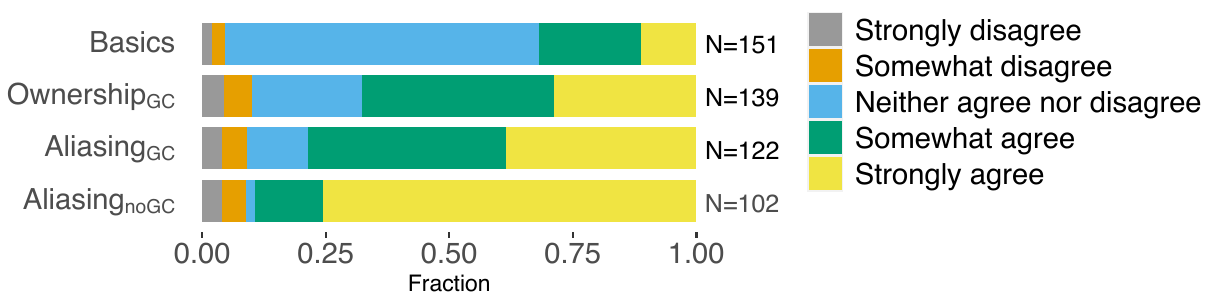}
\caption{Agreement with ``GC in Rust makes writing programs easier'' (Bronze participants only).}
\label{fig:easier-histograms}
\end{figure}

If users feel that GC makes programming easier, then including GC in a language might improve language adoption rates. We were interested in how Bronze participants' opinions of GC changed while using it. Responses to our question about whether GC makes programs easier were on a Likert scale with a 0 to 4 point range, with 4 indicating a strong belief that GC made programming easier. Median scores were 2 after \Basics and 4 after \task{Aliasing}{T}. Thus, after trying the assignment with and without GC, they recognized a strong benefit of GC.  An ordinal regression yielded $p < .001$ (odds ratio 21.4), indicating a significant effect of doing the assignment on beliefs about GC helpfulness. \Cref{fig:easier-histograms} illustrates how participants' beliefs about GC changed over time.

\subsection{Liking Rust: comparison between conditions and over time}
\begin{figure}
\includegraphics[width=\columnwidth]{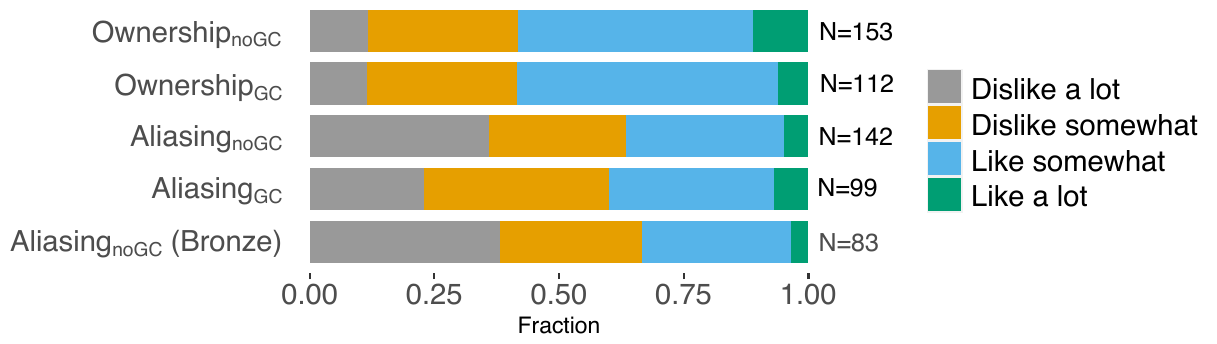}
\caption{Amount participants reported liking Rust on a 4-point Likert scale. \task{Aliasing}{T} (Bronze) indicates that task done by Bronze participants.}
\label{fig:amount.liked.Rust}
\end{figure}

We hypothesized that if Bronze helped participants complete tasks faster, Bronze participants might like Rust better than Traditional participants. Each survey asked: ``How much do you like Rust?'' on a four-point Likert scale (we omitted a neutral option in order to force expressing a preference). \Cref{fig:amount.liked.Rust} shows the responses. We conducted an ordinal regression to assess whether there was a difference in Likert-scale responses across conditions after experiment participants were finished with the project (including those who did not score 100\%). We found $p \approx 1$, indicating no significant effect of GC on liking Rust by the end of the assignment. We also compared responses to this question after \Ownership; the ordinal regression gave $p \approx 1$, also indicating no significant difference.

\subsection{Estimated time and fun in other languages}

We were interested in how usage of Bronze might impact beliefs about time required in \emph{other} languages, since a high estimate of time in another language might correlate with a higher chance of choosing Rust, and since a good experience with GC might lead to higher estimates in languages that do not provide GC. In each survey, we asked participants: ``Suppose you had done this part of the assignment in a different language instead. How much time would it have taken in THAT language compared to using Rust?'' We asked a corresponding question asking about prediction of fun compared to Rust. Responses were on a five-point Likert scale, with an additional ``I'm not familiar enough with this language to judge'' option. We asked about C, C++, Java, and Python. Our exploratory data analysis led us to hypothesize that the relationships were weak, if they could be found at all. The strongest relationship appeared to be with prediction of time in C, and we hypothesized that Bronze participants would estimate longer times in C than the non-Bronze participants, since they would be more aware of the cost of manual memory management. 

To compare predicted time in C across the two conditions (for which we received 864 responses from people familiar enough with C to answer), we used an ordinal regression mixed model. This approach accounted for the fact that we asked the same question of each participant after completing each part of the assignment. We found no significant difference ($p \approx 1$).



\subsection{Correlations among opinions}
\begin{figure}
\includegraphics[width=0.9\columnwidth]{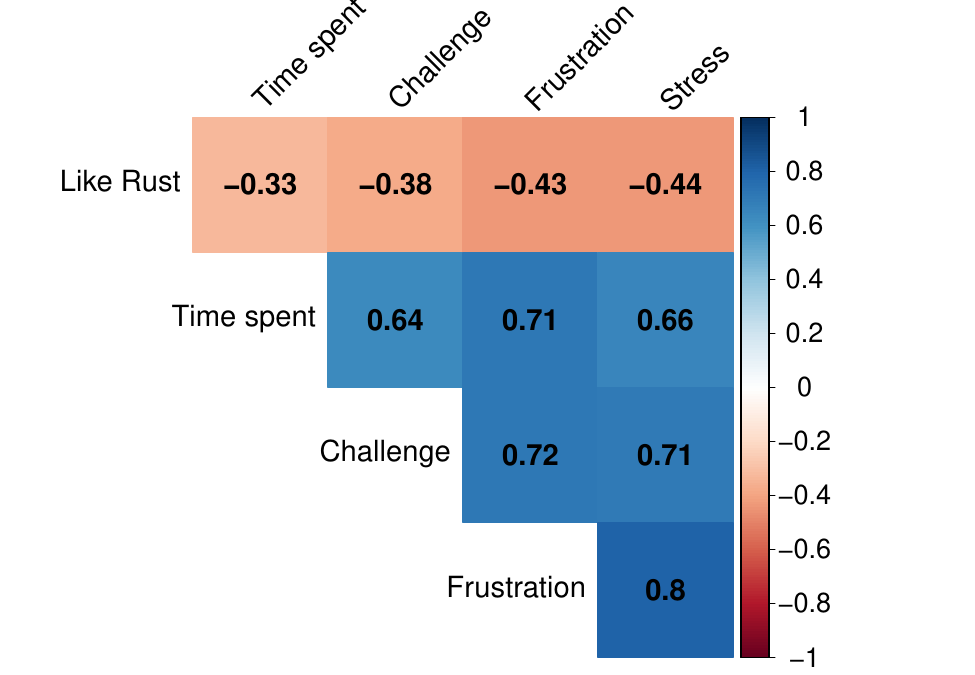}
\caption{Correlations among participants' assessments of their experiences.}
\label{fig:correlations}
\end{figure}

To improve adoption of a language, it might be useful to understand what factors of a programmer's experience correlate with liking a language. After participants completed each part of the assignment, we asked them how much stress, time, intellectual challenge, and frustration they experienced compared to their expectation, as well as how much they liked Rust. We analyzed the correlations among feelings about Rust and other aspects of participants' experiences. \Cref{fig:correlations} shows the pairwise correlations. For the \textit{time} metric, we asked participants to compare the amount of time they spent to the amount of time they expected to spend; we asked this question rather than using the reported times in order to focus on participants' \textit{feelings} about time. Stress, time, intellectual challenge, and frustration appear to be strongly correlated, and assessments of whether participants like Rust are moderately correlated with stress and frustration.

\subsection{Amount learned about Rust}

\begin{figure}
\begin{subfigure}[t]{.49\textwidth}
\includegraphics[width=\columnwidth]{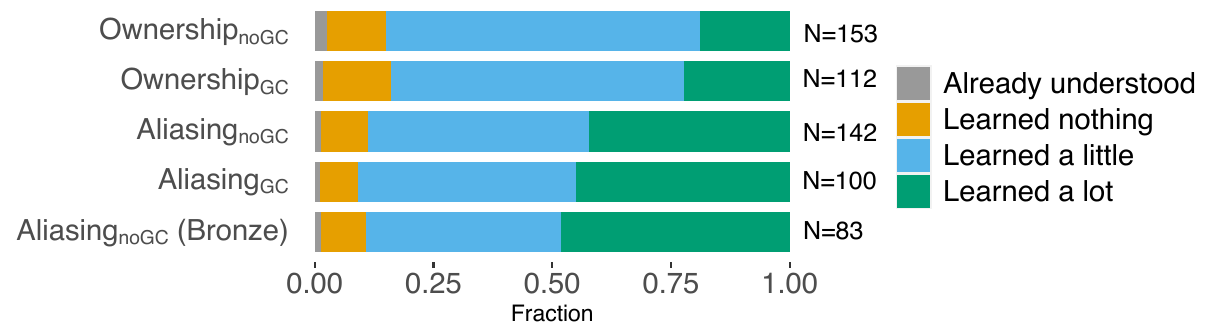}
\caption{Amounts learned about ownership}
\end{subfigure}
\begin{subfigure}[t]{.49\textwidth}
\includegraphics[width=\columnwidth]{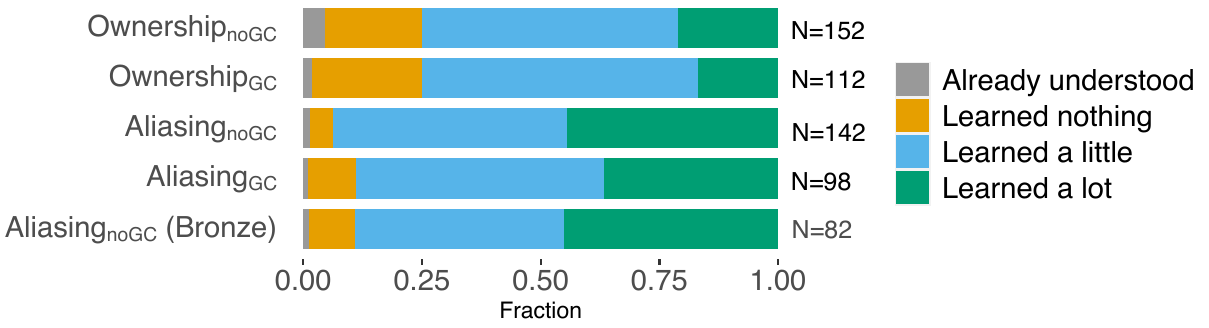}
\caption{Amounts learned about borrowing}
\end{subfigure}
\caption{Amounts learned after each part of the study. \task{Aliasing}{T} (Bronze) indicates that task done by Bronze participants.}
\label{fig:learned.topics}
\end{figure}

After completing each part of the assignment, we asked participants how much they felt they learned about various topics in Rust. \Cref{fig:learned.topics} shows the results for ownership and borrowing. The amount participants reported learning in each part of the assignment does not appear to depend on whether participants used Bronze.

\subsection{Factors influencing grades}

Factors other than usage of Bronze may affect success; we hoped to understand what factors were relevant. In our exploratory data analysis (on 10\% of the data), we used a linear mixed-effects model to evaluate the relationship between methods that participants reported using and their assignment grades. We looked at time spent, sources of help, and choice of developing on their own machine or on a server. Based on this, we hypothesized that course grade and development location might be significant factors. Students who do the work on a department server may have less access to computing resources or less skill at configuring systems. A linear regression found a significant positive correlation with course grade ($p < .001$, adjusted $r^2 \approx .044$, $N = 864$) and that a 1-point increase in course grade corresponded with a $.692$-point increase in assignment grade. Development location was not a significant factor.

Because course grade only explained a small amount of the variance in assignment grade, we did additional exploration with the 10\% withheld samples to identify a stronger predictor of grades, focusing on the whole assignment grade (rather than individual tasks). Returning to the remaining dataset, when predicting score on parts after \Basics from hours spent and course grade, we found that a significant correlation only with grade ($p < .001$, adjusted $r^2 = 0.2465$, $N = 864$). However, when predicting grades for \Aliasing only (both GC and non-GC, with grades re-scaled from 0 to 100\%), we found that time spent and course grade were both significant predictors ($p < .001$ for both factors). The model for \Aliasing explained 20.9\% of the variance in grades ($r^2 \approx .209$).

\section{Qualitative Analysis of Responses}

The surveys included two free-response questions: ``What aspects of this part of the assignment did you find most difficult?'' and ``What should we change about the class to help future students understand the concepts in Rust better?'' Although not all participants tried both conditions, their subjective experiences with only one condition can provide insights about their experiences as well as help understand factors that may relate to their future decisions about programming tools. To analyze the data, we conducted a thematic analysis~\cite{Braun2006:Using}. A single domain expert inductively developed the codebook and coded all responses, in consultation with the research team. Because answers to the two questions overlapped, we analyzed all responses together. We were interested in gleaning insights about language design as well as pedagogy from the responses.

\paragraph{References, lifetimes, ownership among the most challenging aspects}
We received 1,143 comments on challenges that participants faced. 340 pertained to \emph{references}, of which 204 were about \emph{lifetimes}. 116 comments indicated \emph{syntax} had been challenging. 100 mentioned \emph{interior mutability}, 83 \emph{dynamic borrowing}, and 70 \emph{mutability}.

Consistent with earlier reports~\cite{Williams21, Fulton2021:Benefits}, 199 participants reported that ownership was challenging; 76 such reports were about borrowing. One student, after finishing \task{Aliasing}{B}, reported: ``Learning rust ownership is like navigating a maze where the walls are made of asbestos and frustration, and the maze has no exit, and every time you hit a dead end you get an aneurysm and die.'' A Traditional student reported: ``Coding with ownership rules and trying to implement mutability, in general, was just such a headache. It is like someone had combined the worst part of C and Java.''

\task{Aliasing}{T} required designing structures that supported \emph{dynamic borrowing} --- a borrowed reference whose safety was checked dynamically rather than statically. This required understanding the Rust \code{Ref} and \code{RefMut} structures, which dynamically track outstanding borrows of cells. One participant explained the difficulty: ``It took me a long time to figure out the field types with the correct lifetimes\ldots Once I knew what the types were, it took me about an hour or two to figure out the rest of the functions. It was quite confusing and frustrating figuring out how to use Rc and RefCell in the right spots, and my initial approach did not use them nearly as much as my final solution did.''

\paragraph{Error messages can help, but may not aid design or comprehension}
Although Rust's error messages have a reputation for being high-quality~\cite{Fulton2021:Benefits} (one participant wrote ``the Rust compiler pretty much just wrote the program for me''), the compiler cannot give high-level design feedback. One GC participant explained: ``When I was testing my non-GC version, I'd never run into so many errors in my life. When I tried fixing my errors, new ones just came up. I've heard students compare the debugging process for the non-GC version to a never-ending game of whack-a-mole.'' Another reported: ``Translating from GC to no GC was a nightmare. It wasn't as simple as switching \code{GcRef<Turtle>} to \code{Rc<RefCell<Turtle>>} because of the new structs. Figuring out the typing of the structs along with debugging all the mutable references errors that resulted took hours and even some of the TAs couldn't help me.'' Error messages could also lead to working code without teaching the programmer what was wrong: ``Getting weird errors that I still don't understand but just fixed by listening to the compiler.''

Error messages also tend to give local advice that does not necessarily lead the user in the right direction. One participant reported ``error messages were cyclical with things like \code{remove \&} then after removing \code{try adding \&}.'' Three participants said it was possible to fix errors by following instructions without understanding what was going on. A higher-level error management approach, and better integration into an IDE (e.g., with visualizations~\cite{Hill2002:Scalable}), could help.

\paragraph{Garbage collection avoids mutability problems}
We received 103 comments regarding garbage collection. 28 asked us to explain GC more thoroughly. 43 said they had a hard time understanding GC. However, some Bronze participants observed, after finishing the assignment, that GC had been very helpful. One reported: ``After doing this part, I actually realized the power of garbage collection. Using \code{RefCell} and \code{Rc}s to create interior mutability, etc.\ is so hectic. I would always prefer \code{GcRef}! Technically \code{Rc<RefCell<T>>} acts like \code{GcRef} but much more work!'' Another participant said: ``Understanding references, lifetimes, and mutability without garbage collection is very difficult. It is not intuitive or understandable without GC.'' 

Free copying of GC references appeared to be critical to the helpfulness of GC. One participant said: ``The transition from garbage collection to non-gc was rough. I think the garbage collection was as easy as it was because it implemented the Copy trait. My most common error in this project was the one where a certain variable was moved because it was of a type that didn't implement copy. If not for some TA help, I would have been completely lost.''

\paragraph{Students wanted more time and more examples to support learning}
We received 715 comments regarding the course design, assignments, lectures, and recitations. Participants reported that the \Aliasing task was extremely challenging, and that previous parts of the assignment left them unprepared for it. 182 comments asked us to revise the project design or clarify the specifications.

190 comments asked for a longer, more complete treatment of Rust in future versions of the course. 54 asked for an intermediate-level assignment. One participant put the steep learning curve (despite splitting the work over multiple parts) as follows: ``I think having this \ldots project is a bit steep. I felt like I was being thrown into a big steaming wok.''
%
In particular, students requested more or more detailed examples (74), more or continued live coding demonstrations (35), and more discussion sections (31).

\section{Discussion}
\label{sec:discussion}

We discuss key takeaways and lessons learned from our experiment.

\paragraph{Even with GC, Rust learners need to understand ownership, borrowing, and lifetimes}
When designing the experiment, we were concerned that using GC would allow participants to avoid learning about ownership, borrowing, and lifetimes. Because we were hoping Bronze would be an aid to learning traditional Rust, this might have been a problem. However, because of the fundamental way ownership is used in Rust, much of the code required understanding ownership and borrowing even with GC. In this assignment, GC primarily served to aid situations that involved multiple references to values. As \cref{fig:learned.topics} shows, participants reported learning similar amounts about these critical topics across conditions.

\paragraph{Most of the benefit of GC comes from architectural simplification}
Participants reported that the architectural requirements in \task{Aliasing}{T} were extremely challenging; it is likely that the design was a significant contributor to the difference in performance between non-Bronze and Bronze participants. In particular, although Bronze participants only needed to return GC references, Traditional Rust participants participants needed to fill in \code{TurtleRef} and \code{BorrowedTurtle} structures, which required additional design insight. The challenge posed by this design was apparent in the survey responses, in which 100 comments pertained to interior mutability and 83 to dynamic borrowing (almost as many total as the 199 who complained about ownership or borrowing). We conclude that a significant part of the benefit of GC in Rust programs is the architectural simplifications it enables and promotes. 
 
\paragraph{Participants and non-participants were comparable}
We were surprised that students with higher grades were more likely to enroll; we had expected students with lower grades would be more incentivized by the extra credit. Perhaps students with higher grades were more willing to accept the additional work of participating, or perhaps those students care more about even small grade boosts. However, the difference in median course grade between participants and non-participants (90\% vs. 87.0\%) was small enough that we believe that our results likely generalize to the entire class.

\paragraph{Students would have benefitted from more time to complete the assignment}
The students were motivated by grades to complete the assignment, but nearly half of students in both conditions did not finish it (\cref{fig:all.fractions.completed}). Some students reported that they wished the work had been assigned over a longer period of time or earlier in the semester (further from exams). The median participant spent 15 hours on the experiment, which is a bit high for a 12-day homework assignment, and some participants spent significantly more time. If we had allocated more time for the assignment and given it earlier in the semester, perhaps more students would have finished.

\paragraph{Withdrawals were mostly assigned to Bronze}
Of the 41 withdrawals, 40 had been assigned to use Bronze. Eight withdrew within 20 minutes of signing up, suggesting they did not make a serious attempt before switching.  When withdrawing, 24 students reported that they felt the non-GC version would be easier, perhaps because the GC version required completing an additional part of the assignment. Six said they didn't understand GC well enough or that it was poorly documented. Eleven students gave no explanation.

We suspect the withdrawal rate would have been lower if we could have convinced participants that they were likely to spend the same amount of time total regardless of which condition they picked. Of course, this approach may not be practical in situations where the expected times are not yet known. Future experiments could provide incentives that depend on time spent; our design incentivized unbiased time reporting.

\paragraph{Encouraging adoption of safer languages by reducing stress}
The apparent decrease over time in how much students like Rust (\cref{fig:amount.liked.Rust}) suggests that if one wants to encourage Rust adoption, changes to the assignment design are needed. It would appear that the \Aliasing task had the largest influence on (dis)liking Rust. 

It might be surprising that Bronze participants did not like Rust any better than Traditional Rust participants after \task{Aliasing}{B}, which Bronze participants completed in about a third of the time of Traditional Rust participants. We observed that stress and frustration correlate more closely with liking Rust than whether the task took longer than participants expected. Expectancy-value theory~\cite{Wigfield2000:Expectancy-Value} suggests that an expectation of success contributes to people's motivation in doing tasks. The theory might suggest that if we want to encourage adoption of Rust and other safe languages, it is more important to provide users with a consistent feeling of progress rather than focusing on minimizing total task completion time. Predictability is beneficial for practicing software engineers as well as for students, and language adopters must first learn the language before using it in a project, so this emphasis is valuable in practice as well as in education.

\section{Related Work}

Empirical studies have been used to study several programming language design questions, such as static typing~\cite{Hanenberg2014:Empirical}, lambdas in C++~\cite{Uesbeck2016:Empirical}, immutability features~\cite{Coblenz2017:Glacier}, and typestate~\cite{Coblenz2020:Can}. Qualitative studies have also been used to understand what factors contribute to users' perceptions of languages~\cite{Coblenz2021:PLIERS}. This is the first empirical study of the usability of garbage collection of which we are aware.

Some work investigated how Rust is used in the wild. Astrauskas et al. investigated the use of the \code{unsafe} keyword~\cite{Astrauskas2020:How}, finding that much code relies on \code{unsafe}. Fulton et al.~\cite{Fulton2021:Benefits} conducted a survey and interviews of Rust programmers to understand their motivations for adopting or not adopting Rust, finding that programmers are motivated by the safety benefits but concerned about the learning curve and challenges in hiring experienced Rust programmers.

Other GCs for Rust include rust-gc~\cite{rust-gc}, Shifgrethor~\cite{Shifgrethor}, and Josephine~\cite{Josephine}, which require manual specification of roots. Josephine is for implementation of JavaScript in Rust. Shredder~\cite{Shredder} supports concurrency, unlike Bronze, but accessing a GC object requires obtaining a guard to prevent concurrent access. It manages roots automatically by keeping a global list of all allocations. As a result, references do not implement the \code{Copy} trait and therefore cannot be copied freely as they can in Bronze.

Meyerovich investigated programming language adoption~\cite{Meyerovich2013:Empirical}; developers reported preferring more-expressive languages. Because adding optional garbage collection allows developers to express different kinds of aliasing structures than does Rust alone, adding GC to Rust might make it more likely to be adopted. Zeng and Crichton~\cite{Zeng2018:Identifying} investigated forum posts about Rust adoption, hypothesizing that adoption barriers for Rust included poor tool publicity, difficulty solving complex aliasing problems, and integration challenges with existing contexts. GCs, such as Bronze, may help make it easier for programmers to solve complex aliasing problems. 

RustViz~\cite{Luo2020:Rustviz} is a visualization tool that may help programmers learn Rust ownership semantics.

Cyclone~\cite{GrossmanMJHWC02} integrated region-based memory management~\cite{tofte97regions}, including an optional garbage collector, into a safe dialect of C\@. Later extensions included support for ownership and borrowing~\cite{swamy05experience}. Case studies found that similar performance could be obtained if the GC was used judiciously, but that GC can have a significant performance cost if used globally.

\section{Conclusions and Future Work}

We developed Bronze, a new library-based GC whose goal is to ease the learning and use of Rust. We carried out a randomized, controlled trial of Bronze that showed that it can significantly alleviate some of the challenges posed by the Rust aliasing restrictions for Rust beginners: Bronze participants completed a task that required a complex aliasing structure in about a third as much time as traditional Rust participants. GC may enable Rust programmers, particularly beginners, to complete tasks in much less time. 

We hope to extend the collector to trace arbitrary objects that may transitively contain references to GC objects. We would like to investigate the impact of using GC not just for complex aliasing scenarios, but to mitigate the impact of ownership in general; perhaps doing so could flatten the learning curve and help users feel more positively about Rust. We would also like to relax the single-thread assumption in the collector, which would likely entail changing from the \textit{shadow stack} GC\footnote{https://llvm.org/docs/GarbageCollection.html\#the-shadow-stack-gc} that is currently in use.

The experiment is the first (to our knowledge) evaluating the usability benefits of GC. We focused on using GC to relax aliasing restrictions, but GC can also be used to avoid the challenges of reference counting and manual memory allocation. In the future, library-based garbage collection could be used to evaluate the usability tradeoffs of garbage collection in other contexts as well.

\begin{acks}
	We appreciate Dan Votipka's feedback on drafts of this paper, as well as the feedback from the anonymous reviewers.  We thank Alan Jeffrey, Manish Goregaokar, Felix Klock, and Philip Reames for insightful discussions regarding garbage collection and Rust. We also thank those who piloted the experimental materials, including Michael Dong, Aaron Eline, and the course TAs, who supported the experiment while it was running. Finally, we appreciate the helpful design suggestions and guidelines from our IRB. This material is based upon work supported by the National Science Foundation under Grant No. 1801545. Any opinions, findings, and conclusions or recommendations expressed in this material are those of the author(s) and do not necessarily reflect the views of the National Science Foundation.
\end{acks}

\newpage

\bibliographystyle{ACM-Reference-Format}
\bibliography{bronze}

\end{sloppypar}

\end{document}